\documentclass[twocolumn,10pt,aps,prb,reprint, superscriptaddress, showpacs, showkeys]{revtex4-1}
\usepackage{amsfonts}
\usepackage{amsmath}
\usepackage{amssymb}
\usepackage{graphicx}
\usepackage{color}

\begin{document}

\title{Superparamagnetic Relaxation Driven by Colored Noise}

\author{J. G. McHugh}
\affiliation{Department of Physics, The University of York, York, YO10 5DD, UK}
\author{R. W. Chantrell}
\affiliation{Department of Physics, The University of York, York, YO10 5DD, UK}
\author{I. Klik}
\affiliation{Department of
Physics, National Taiwan University, Taipei, Taiwan}
\author{C. R. Chang}
\affiliation{Department of
Physics, National Taiwan University, Taipei, Taiwan}
\vspace*{15mm}

\begin{abstract}
A theoretical investigation of magnetic relaxation processes in single domain particles driven by colored noise is presented. Two approaches are considered; the Landau-Lifshitz-Miyazaki-Seki equation, which is a Langevin dynamics model based on the introduction of an Ornstein-Uhlenbeck correlated noise into the Landau-Lifshitz-Gilbert equation and a Generalized Master Equation approach whereby the ordinary
Master Equation is modified through the introduction of an explicit memory kernel.
It is found that colored noise is likely to become important for high anisotropy materials where the characteristic system time, 
in this case the inverse Larmor precession frequency, becomes comparable to the correlation time. When the escape time is much 
longer than the correlation time, the relaxation profile of the spin has a similar exponential form to the ordinary LLG equation, while for
low barrier heights and intermediate damping, for which the correlation time is a sizable fraction of the escape time, 
an unusual bi-exponential decay is predicted as a characteristic of colored noise. At very high damping and correlation times,
the time profile of the spins exhibits a more complicated, noisy trajectory.

\end{abstract}

\maketitle
\section{Introduction}
Thermally-activated magnetization reversal over an anisotropic energy barrier is the driving force for switching in magnetic materials. Theoretical understanding  was first developed by N\'{e}el\cite{Neel} based on the transition state theory (TST) leading to an Arrhenius-like relaxation time proportional to $\exp(E_B/k_B T)$ where $E_B$ is the energy barrier, $k_B$ the Boltzmann constant and $T$ the temperature. Brown \cite{Brown} provided further insight through
the construction of
the Langevin equation for the problem  by the introduction of
white-noise fields into the Landau-Lifshitz equation with Gilbert
damping, leading to the stochastic Landau-Lifshitz-Gilbert (LLG) equation, An expression for the relaxation time of thermally-driven escape over the energy barrier is then found through the
lowest eigenvalue of the corresponding 
Fokker-Planck equation (FPE) governing the time-evolution of the probability density
function of the magnetization orientation.

The route to the Arrhenius-like relaxation time expression is one of two directions leading from the Langevin equation. The second, Langevin Dynamics (LD) approach is the direct numerical solution of the Langevin equation \cite{Berkov,Andreas,Garcia,Berkov2}. There is a natural separation of timescales, with LD used for high frequency applications such as magnetic recording and the Arrhenius-like relaxation time used for slow dynamic behavior arising from thermal activation over energy barriers. The two approaches have been compared by Kalmykov et. al.,\cite{reversal-1} who calculated escape times for both cases giving excellent agreement  for the variation of escape time with damping constant and demonstrating the importance of starting the LD calculations from the correct thermal equilibrium distribution within the energy minimum. 

The LLG equation for a single spin takes the well-known form
\begin{equation}
\frac{d\mathbf{S}}{dt} = - \frac{\gamma}{1 + \alpha^2} \big( \mathbf{S} \times \mathbf{H} + \alpha \mathbf{S} \times (\mathbf{S} \times \mathbf{H})\big) ,
\end{equation}
where $\alpha$ is the phenomenological damping constant, $\gamma = 1.76 \time 10^{11} T^{-1} s^{-1}$ and $\mathbf{S}$ is a unit
vector in the direction of the spin, $\mathbf{S} = \boldsymbol{\mu}/\mu_s$. The local magnetic field, $\mathbf{H}$, 
is derived from the first derivative of the spin Hamiltonian $\mathcal{H}$ with  respect to the spin degree of freedom,
\begin{equation}
\mathbf{H} = - \frac{1}{\mu_s} \frac{\partial \mathcal{H}}{\partial \mathbf{S}}.
\end{equation}

Thermal fluctuations are necessary to incorporate the deviations of a particular spin from the average
trajectory. This is done via the formal inclusion of random
fields in the LLG equation.
In order to realize the Fluctuation-Dissipation theorem for this system, these thermal fields
must also be proportional to the same phenomenological damping constant, $\alpha$ that occurs in the damping. The moments of the thermal field are then given by
\begin{equation}
\langle H_{th,i}(t) \rangle = 0
\end{equation}
\begin{equation}
\langle H_{th,i}(t) H_{th,j}(t') \rangle = \frac{2 \alpha k_B T}{\gamma \mu_s} \delta(t-t') \delta_{ij}
\label{Langevin}
\end{equation}
where $i,j$ label the spin components.

In all numerical simulations, we interpret the stochastic equation in the Stratonovich sense and employ the Heun method
An implicit assumption of this approach is the presence of white noise, which exists in the zero correlation
time limit for some physical noise process with a well-defined correlation time. 
Such a colored noise may be implemented for a magnetic system through the use of the Landau-Lifshitz-Miyazaki-Seki
pair of Langevin equations, which take the form
\begin{equation}
\frac{d\mathbf{S}}{dt} = \gamma \mathbf{S} \times \big(\mathbf{H} + \boldsymbol{\eta} \big),
\label{LLMS-1}
\end{equation}
\begin{equation}
\frac{d \boldsymbol{\eta}}{dt} = -\frac{1}{\tau_c} (\boldsymbol{\eta} - \chi \mathbf{S}) + \mathbf{R},
\label{LLMS-2}
\end{equation}
where $\tau_c$ is the correlation time and $\chi$ is a spin-bath coupling which is related to the
phenomenological damping parameter as $\alpha = \gamma \chi \tau_c$
in the limit of small correlation times.
The autocorrelation of the white noise field, $\mathbf{R}$, is
given by
\begin{equation}
\langle R_i(t) R_j(t') \rangle = \frac{2 \chi k_B T}{\tau_c \mu_s} \delta_{ij} \delta(t-t').
\end{equation}
This pair of Langevin equations leads to a frequency-dependent damping
of the spin together with an exponentially correlated noise term in the spin-only space,
\begin{equation}
\langle \hat{\eta}_i(t) \hat{\eta}_j(t') \rangle = \frac{\chi k_B T}{\mu_s} e^{\frac{-(t-t')}{\tau_c}}\delta_{ij} = \frac{\chi k_B T}{\mu_s} K(t - t') \delta_{ij}
\end{equation}
where $K(t) = \exp{\frac{-(t-t')}{\tau_c}}$ is the exponential memory kernel.
For completeness, additional background on the LLMS Langevin equation and colored noise is included in Appendix A.

An alternative approach to the Langevin equation is the  discrete orientation approximation, whereby, in the limit of large barriers, the detailed dynamics are
replaced by phenomenological rate equations describing
transitions between the minima of the magnetic potential. We may augment this description by the
introduction of a memory kernel into the rates, thus
replacing the master equation description with a generalized master equation which explicitly incorporates the 
retardation effect into the rate equations.

Here we investigate the introduction of colored noise into the calculation of escape rates. This
leads to significant effects for materials with large magnetocrystalline anisotropy energies, 
including the prediction of bi-exponential behavior at
intermediate damping, when the characteristic time of the relaxation process becomes comparable to the heat bath correlation time. 
The paper is organized as follows. We first outline thermally activated escape times for single nanoparticles, followed by an introduction of colored noise into the Langevin formalism via the LLMS equations.
We then derive the relaxation profile from the non-Markovian generalized extension of the rate equation,
followed by a systematic investigation of the effects of the barrier height and correlation times on the relaxation profile 
from LLMS simulations.

\subsection{Thermally-Assisted Magnetization Reversal}
We will investigate here the effect that colored noise has on the dynamics of the
thermal escape problem for a magnetic nanoparticle. The
spin Hamiltonian of the system contains both an applied field and anisotropy term,
taking the form
\begin{equation}
\mathcal{H} = - KV \mathbf{S}_z^2 - \mu_s \vec{\mathbf{H}} \cdot \mathbf{S},
\end{equation}
where $K$ is the anisotropy constant and $V$ is the particle volume. 
For the escape problem we have a spin energy potential of the form
\begin{eqnarray}
V(\theta, \phi) & = & \sigma \beta^{-1} \big( \sin^2 \theta -  2 h (\cos \psi \cos \theta \\ \nonumber
& & + \sin \psi \sin \theta \cos \phi)  \big),
\end{eqnarray}
where $\theta$, $\phi$ are respectively the polar and azimuthal components of the spin in spherical coordinates, 
$\sigma = KV/k_B T$ is the reduced barrier height parameter, $h = H/2 \sigma$ is the reduced field,
$\beta = (k_B T)^{-1}$ and $\psi$ is the angle between
the easy-axis and the applied field. 
This potential has a bistable character
under the condition than the
critical applied field value, $h < h_c(\psi) = ((cos^{2/3} \psi 
+ sin^{2/3} \psi)^{-3/2}$ \cite{nowak-review}, in which case there are local
and global minima in the north and south polar regions, with an equatorial saddle point
between them. We are then interested in the calculation of the characteristic
escape time of a spin initialized in one such minimum.

For the special case of aligned field and easy axis, for which $\psi = 0$ the potential is
\begin{equation}
V(\theta) = \sigma \beta^{-1} \big(  \sin^2 \theta - 2h \cos \theta \big).
\end{equation}
In this case the
escape time takes the Arrhenius form, where the barrier energy, $E_B$,
is  proportional to the anisotropy energy, leading to an escape time
\begin{equation}
\tau \propto f_0^{-1} e^{KV/k_BT}
\end{equation}
where $f_0$ is the attempt frequency, the frequency of Larmor gyromagnetic precession at the bottom of the well.

We investigate the escape time in the colored and white noise cases through repeated numerical
integration of the Langevin equations for a spin initialized in a potential minimum. An important
consideration for such simulations is the choice of initial and switching condition for the spin. 
We will initialize the spins with the Boltzmann distribution at the bottom of the well in order 
to avoid inconsistencies at low damping, while the switching condition is chosen such that $S_z < -0.5$,
with the spin initialized in the positive $z$-direction, so that the spin is sufficiently deep in the
well such that it has escaped.

\section{LLMS Escape times \& Colored Noise}
\subsection{System time $\tau_s$ vs. $\tau_c$ characteristic bath time.}
For the uniaxial escape problem the external field in the LLMS will consist of an external applied
part and an anisotropy contribution
\begin{equation}
\mathbf{H} = \mathbf{H}_a + \mathbf{H}_0
\end{equation}
the magnitude of the anisotropic contribution depends on the orientation of the spin and is given by 
$\mathbf{H}_a = \frac{2 k_u}{\mu_s} \vec{\mathbf{S}}_z \cdot \vec{\mathbf{z}} = H_k \vec{\mathbf{S}}_z \cdot \vec{\mathbf{z}} $ 
where $\vec{\mathbf{z}}$ is the direction of easy magnetization and $k_u$ is the anisotropy energy. To
gain intuition into the relevant
timescales for the relaxation problem, we will assume the
uniaxial case in the following, where the external field is applied 
along the same direction as the easy axis, such that both fields only have components in the $z$-direction. 

We note that the anisotropic field contribution varies with the projection of the
spin on to the easy-axis as
\begin{equation}
\mathbf{H}_a = H_k (\mathbf{S} \cdot \mathbf{z}) \vec{\mathbf{z}}= (H_k \cos \theta) \vec{\mathbf{z}},
\end{equation}
The {\it largest field magnitude} and consequently the fastest timescale of the problem
is set by the value for which the anisotropic field contribution is at its largest, which is when the spin and the easy-axis precisely coalign. For any other orientation, the field will be smaller and the timescale of oscillation hence slower. 
We may then take the spin-only Langevin equation,
\begin{equation}
\frac{d\mathbf{S}}{dt} = \gamma \mathbf{S}(t) \times 
\big((H_k \cos{(\theta)})\vec{\mathbf{z}} + \bar{\boldsymbol{\eta}} - \chi \int_{-\infty}^t dt' K(t-t') \frac{d\mathbf{S}(t')}{dt'} \big),
\end{equation}
and proceed to scale this equation by the maximum anisotropy field value.
Defining the system time for the spin as $\tau_s = (\gamma H_k)^{-1}$ then
\begin{eqnarray}
\frac{d\mathbf{S}}{dt} & = & \frac{1}{\tau_s} \mathbf{S}(t) \times \big(\cos{(\theta)} \vec{\mathbf{z}} + H_k^{-1} \bar{\boldsymbol{\eta}} \nonumber \\
& - & H_k^{-1} \chi \int_{-\infty}^t dt' K(t-t') \frac{d\mathbf{S}(t')}{dt'}\big) .
\end{eqnarray}

\begin{figure}[ht]
\begin{center}
\includegraphics[width=0.5 \textwidth, height=\textheight,keepaspectratio]{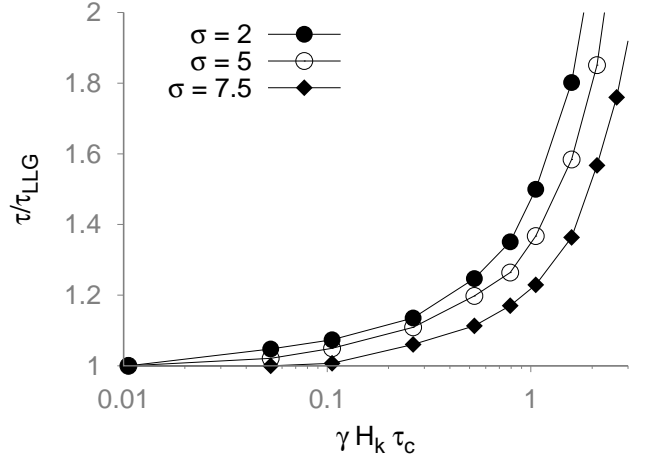}
\end{center}
\caption{Escape time, normalized to the uncorrelated LLG 
escape time vs correlation time, from LLMS simulations for
a Co nanoparticle with $\alpha = 0.05$ and
different reduced barrier heights, $\sigma$, }
\label{fig1}
\end{figure}

We may scale the time variable in the Langevin equation so that the system time is removed by taking $\zeta = \tau_s t$. Then
we have
\begin{eqnarray}
\frac{d\mathbf{S}}{d\zeta} & = & \mathbf{S}(\zeta) \times \cos{(\theta)} \vec{\mathbf{z}}  +  
		\mathbf{S}(\zeta) \times \bigg( H_k^{-1}\bar{\boldsymbol{\eta}}(\zeta) \nonumber \\
		& + & H_k^{-1} \chi \int_{-\infty}^{\zeta'} d\zeta' e^{-(\zeta-\zeta') \frac{\tau_s}{\tau_c}} \frac{d\mathbf{S}(\zeta')}{d\zeta'}\bigg)
\end{eqnarray}
The autocorrelation of the noise is similarly transformed to become
\begin{equation}
\langle \bar{\boldsymbol{\eta}}(\zeta) \bar{\boldsymbol{\eta}}(\zeta') \rangle = \frac{\tau_s}{\tau_c} \bar{D} e^{-\zeta-\zeta') \frac{\tau_s}{\tau_c}}
= \frac{\bar{D}}{\tau} e^{-(\zeta-\zeta')/\tau}
\end{equation}
where $\tau_s/\tau_c = \tau$ and $\bar{D} = D/\tau_s = \chi \tau k_B T/\mu_s$. We can then write the coupling
as $\bar{\chi} = \chi/H_k$, and absorb the $H_k$ factor into the diffusion constant for the thermal field. Since
the thermal fields are given by $\bar{\boldsymbol{\eta}}(\zeta) = \frac{\sqrt{2D}}{\tau} \int_{-\infty}^\zeta K(\zeta-\zeta') \Gamma(\zeta')$, the
diffusion constant becomes
\begin{equation}
\bar{D} = \frac{\chi \tau k_B T}{\mu_s H_k^2} = \frac{\bar{\chi} \tau k_B T}{2 k_u} = \frac{\bar{\chi} \tau}{2 \sigma}
\end{equation}
where $\sigma = k_u/k_B T$.
The final expression for the Langevin equation is then
\begin{equation}
\frac{d\mathbf{S}(\zeta)}{dt} = \mathbf{S}(\zeta) \times \big(\cos{(\theta)} \vec{\mathbf{z}} + 
\bar{\boldsymbol{\eta}} - \bar{\chi} \int_{-\infty}^\zeta d\zeta' K(\zeta-\zeta') \frac{d\mathbf{S}(\zeta')}{d\zeta'}\big).
\end{equation}

In the case that $\tau \ll 1$ and $\tau_c \ll \tau_s$, the memory kernels appearing 
in the noise and damping terms are reduced to delta functions and the white noise behavior is restored. Additionally the bath coupling
and the strength of the thermal fluctuations are reduced by the anisotropy field, so that in the event
of a very large anisotropy the precessional dynamics of the spin dominate the thermal and damping parts.
We then conclude that the condition $\tau_c \gtrsim (\gamma H_k)^{-1}$ 
dictates
whether the effect of correlations are relevant in the system dynamics in the high barrier limit.

\begin{figure}[ht]
\begin{center}
\includegraphics[width=0.5 \textwidth, height=\textheight,keepaspectratio]{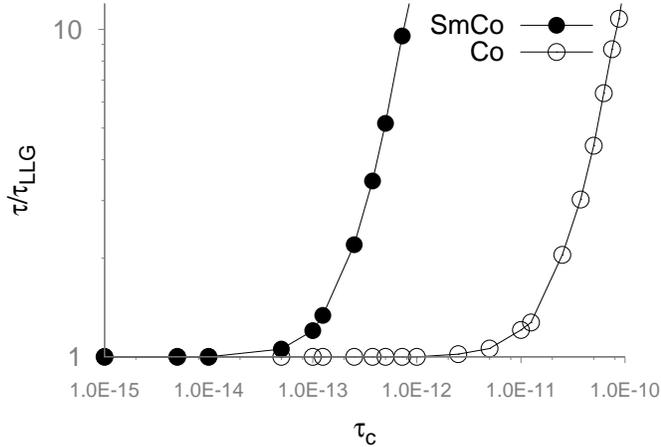}
\end{center}
\caption{Comparison of simulation results for systems with parameters chosen to be similar to 
SmCo\textsubscript{5} and Co nanoparticles, respectively, for large reduced barriers $\sigma = 13.5$, and
a fixed $\alpha = 0.05$. The higher anisotropy SmCo\textsubscript{5} exhibits departure from LLG behavior at 
smaller correlation times.}
\label{fig2}
\end{figure}

This prediction is borne out in numerical simulations of the LLMS equation.
Figure \ref{fig1} depicts the escape time calculated using the LLMS model for a Co nanoparticle of volume $V = 8 \times 10^{-27} m^3$, 
with anisotropy energy $KV = 1.12 \times 10^{21} J$, and a magnetic moment
$\mu_s = 1.12 \times 10^{-20} J/T$.where the correlation time is normalized
by the inverse of the Larmor precession frequency, and the escape time in the LLMS is normalized
by the escape time calculated from the Markovian LLG equation.
The escape rate departs from the LLG escape rate only once the correlation time is
some significant fraction of the Larmor time, and for increasing barrier height the correlation time
must be a larger fraction of the gyromagnetic precession before the escape rate departs from the 
LLG prediction.

Figure \ref{fig2} shows a comparison of the 
escape time for the Co nanoparticle and
a SmCo\textsubscript{5} nanoparticle of the same volume. 
The SmCo\textsubscript{5} material parameters are taken to be $\mu_s = 6.4 \times 10^{-18} J/T$,
and anisotropy $KV = 2.16 \times 10^{-16}$, a much higher anisotropy energy density than Co.
This higher anisotropy gives the nanoparticle a faster system time, which causes the LLMS
to depart from the LLG for smaller bath correlation times, $\tau_c$, on the order of $50-100
fs$ for the SmCo\textsubscript{5} particle, while it is approximately
$1 ps$ for the Co nanoparticle. The fact that the system time is inversely proportional to the 
magnitude of the anisotropy field is exhibited in the simulations by the difference between LLMS
and LLG escape rates at smaller values of the bath correlation time for the material with
higher magnetic anisotropy.

\subsection{Arrhenius Behavior}
\begin{figure}[ht]
\begin{center}
\includegraphics[width=0.31 \textwidth, height=\textheight,keepaspectratio]{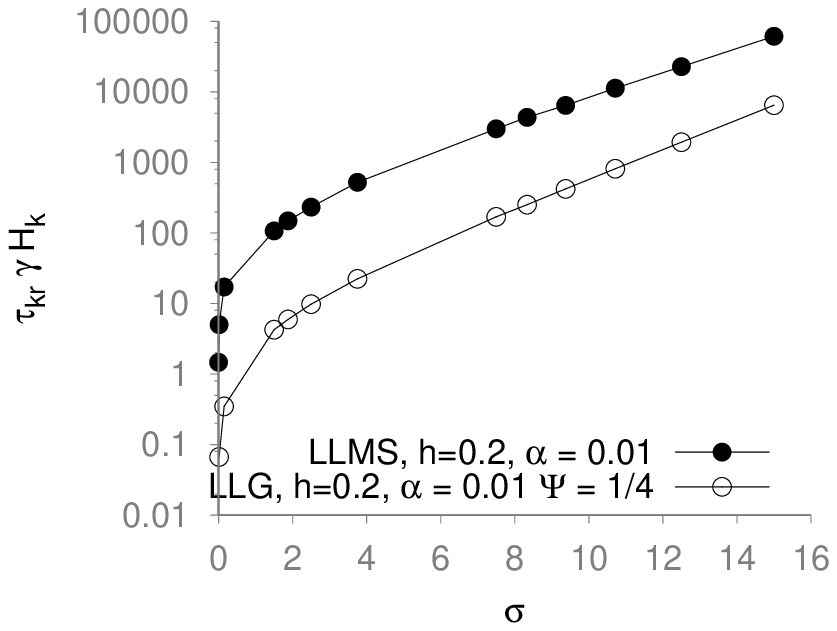}
\includegraphics[width=0.31 \textwidth, height=\textheight,keepaspectratio]{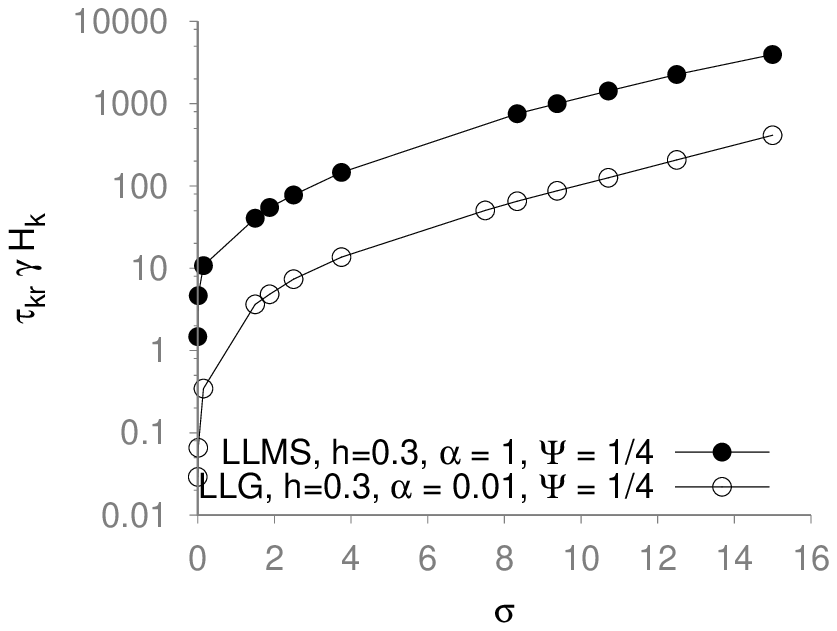}
\includegraphics[width=0.31 \textwidth, height=\textheight,keepaspectratio]{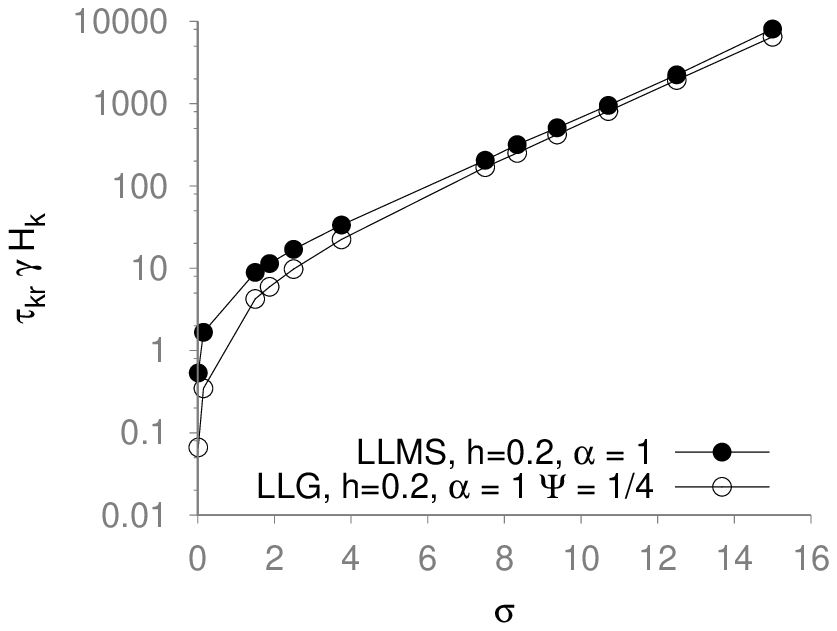}
\includegraphics[width=0.31 \textwidth, height=\textheight,keepaspectratio]{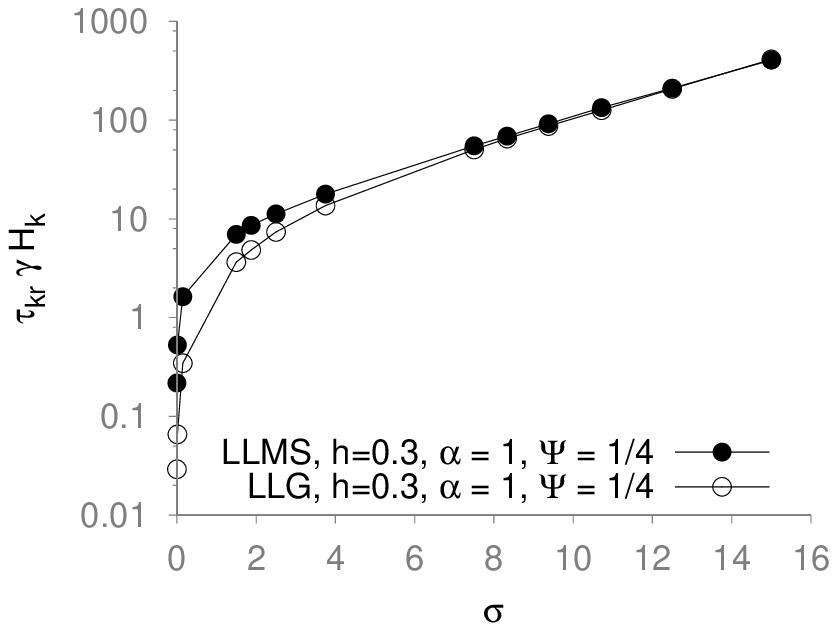}
\end{center}
\caption{Escape time, $\tau \gamma H_k$ vs reduced barrier height, $\sigma$, from LLMS
and LLG simulations,
for different values of the applied field $h = \mu_s H/\sigma$
and damping, $\alpha$, with a fixed angle of $\Psi = \pi/4$ between
the applied field and the easy axis of magnetization.
{\bf 1:} Low damping, $h = 0.2$, {\bf 2:} Low damping, $h = 0.3$.
{\bf 3:} High damping, $h = 0.2$, {\bf 4:} High damping, $h = 0.3$.}
\label{arrhenius}
\end{figure}

Crucially, it is found that the Arrhenius behavior of the
escape rate is recovered from LLMS simulations in the limit of large barrier height. In figure \ref{arrhenius} we show the temperature-
dependence of the escape time vs reduced barrier height.

In the  high damping case, we see that the escape rates begin to converge as the temperature tends towards zero. As the escape time between the wells becomes much longer than the bath correlation time, the detailed dynamics of the spin within the well becomes less relevant.

At low damping, the LLMS and LLG appear not to converge even at the larger
barrier heights considered here. We attribute this difference to the difference in damping regimes and the physically distinct mechanisms involved in the escape process between the two regimes. Escape
at high damping is mediated by thermal fluctuations, which liberate the bound spin. In the limit of vanishing temperature the infrequency of thermal oscillations of sufficient energy dominate the escape behavior and the escape rates converge.

In contrast, the energy-controlled diffusion regime is characterized by the almost-free precessional motion of the  spin in the well. In the highly correlated case, the simple damping is replaced with a frequency-dependent damping, an effect which increases the overall effective damping. In the limit $T \rightarrow 0$, this inhibits the escape rate between the wells by decreasing the rate at which the spin is able to attain a trajectory with sufficient escape energy.

\section{Rate Equations for Thermally-Activated Magnetization Reversal}

\subsection{Master Equation}

The master equation is a phenomenological set of
first-order differential rate equations for a multi-level system, 
which takes the form
\begin{equation}
\frac{dn_i}{dt} = \Gamma_{ij}(t)n_j(t),
\label{ME}
\end{equation}
where $n_i$ is a probability vector representing the probability
that the system is in one of a discrete set of states, and 
$i,j$ label those discrete states, while the matrix of coefficients
$\Gamma_{i,j}$ dictates the transition rate from state $i$ to 
the state $j$ of the system. 

The dynamics of the thermally-assisted escape problem in a magnetic
system may be approximated by such a master equation under the condition that the energy barrier
is large compared to the thermal energy, $\sigma > 1$, but not too large such that it would inhibit inter-well
transitions. This approximation to the Langevin dynamics is 
called the {\it discrete orientation approximation}. The spin orientations are assumed to be restricted only to the $2$ minima of the potential energy dictated by the spin Hamiltonian.
The time evolution of the occupation of each state follows from
Eq. \ref{ME}, where $i,j = 1,2 $. The transition matrix elements
follow from the applied field, anisotropy and temperature. In particular, we will assume a fixed applied field, such that the 
transition rates are constant in time and the matrix takes the form
\begin{equation}
\Gamma_{ij} = \left( \begin{array}{ccc}
- \kappa_{12} & \kappa_{21}  \\
\kappa_{12} & - \kappa_{21}  \end{array} \right),
\end{equation}
In the uniaxial
case these rates are given by $\kappa_{1 \rightarrow 2} = \kappa_{12} = f_0 \exp({- \sigma (1+h)^2})$ and 
$ \kappa_{2 \rightarrow 1} = \kappa_{21} = f_0 \exp({- \sigma (1-h)^2})$, where $\sigma$ and $h$ are the reduced barrier height and 
applied field, respectively.  The time evolution of the
population of the state $n_1$ is then explicitly given by
\begin{equation}
\frac{dn_1}{dt} = -\kappa_{12}n_1 + \kappa_{21}n_2 = (\kappa_{12} + \kappa_{21})n_1 + \kappa_{21} \label{me_n}.
\end{equation}

The time-evolution of the magnetization follows from the 
individual rates for the two wells, where the magnetization is given by $m(t) = n_1(t) - n_2(t)$ and is subject to the normalization condition $n_1(t) + n_2(t) = 1$. The differential
equation for the magnetization is then
\begin{equation}
\frac{dm}{dt} = - \Gamma_1 m(t) - \Gamma_2,
\end{equation}
where $\Gamma_1  = \kappa_{12} + \kappa_{21}$ and $\Gamma_2  = \kappa_{12}-\kappa_{21}$. This is the same form as the rate for the 
individual wells, Eq. \ref{me_n}. For an initial magnetization
$m_0 = n_1(t=0) - n_2(t=0)$, the magnetization as a function
of time is a simple exponential,
\begin{equation}
m(t)  = \frac{e^{-\Gamma_1 t}(\Gamma_1  m_0 + \Gamma_2 )}{\Gamma_1 } - \frac{\Gamma_2}{\Gamma_1 },
\label{m_v_t}
\end{equation}
which tends to the value  
\begin{equation}
\frac{-\Gamma_1 }{\Gamma_2 } = \frac{\kappa_{21} -\kappa_{12}  }{\kappa_{12}+\kappa_{21}}.
\end{equation}
In the long-time limit, the steady state magnetization corresponding to the difference in the transition rates between the wells,
if $\kappa_{2 \rightarrow 1} > \kappa_{1 \rightarrow 2}$, the transition rate into
well $1$ is greater than the rate out, and we have a positive magnetization, as expected.





\subsection{Generalized Master Equation}
The non-Markovian extension of the master equation formalism
is what is called a generalized master equation. Under this model,
the set of $i \times j$ rates represented in the transition matrix in Eq. \ref{ME} are promoted to a set of $i \times j$ memory
kernels for the transitions between the wells $i, j$, 
replacing the set of first-order differential equations with a 
set of integro-differential equations for the population
of each well,
\begin{equation}
\frac{dn_i}{dt} = \int_0^{\infty} M_{ij}(t-\tau)n(\tau)d \tau.
\end{equation}

We will consider the simplified case
\begin{equation}
M_{ij}(t) = \frac{e^{-t/\Theta}}{\Theta} A_{ij} = K(t)\Gamma_{ij},
\end{equation}
where $\Gamma_{ij}$ are the same constant transition rates considered in the Markovian master equations, now modified
by a simple exponential kernel over the recent
population of the well. The integro-differential expression for the
magnetization then becomes
\begin{equation}
\frac{dm}{dt} = - \Gamma_1 \int_0^{\infty}K(t-\tau) m(\tau) d \tau - \Gamma_2 \int_0^{\infty}K(t-\tau) d \tau.
\end{equation} 
Where we note that for the exponential kernel, $K(t) = \frac{e^{-t/\Theta}}{\Theta}$, 
the uncorrelated form of the master equation is recovered in the limit of vanishing correlation time,
$\lim_{\Theta \rightarrow 0} K(t) = \delta(t)$.

The Laplace transform of this equation is
\begin{equation}
\omega m(\omega) - m_0 = - \Gamma_1 K(\omega)m(\omega) - \frac{\Gamma_2}{\omega}K(\omega),
\end{equation} 
where $K(\omega) = \mathcal{L}(K(t))$ is the Laplace
transform of the memory kernel,
\begin{equation}
K(\omega) = \frac{\Theta^{-1}}{\omega + \Theta^{-1}} = \frac{1}{1 + \Theta \omega},
\end{equation}
we then have
\begin{equation}
m(\omega) = \frac{-\frac{\Gamma_2}{\omega}K(\omega) + m_0}{\omega + \Gamma_1 K(\omega)}.
\end{equation}
After inserting the expression for the Laplace transform of the kernel we find
\begin{equation}
m(\omega) = \frac{-\frac{\Gamma_2}{\omega} + m_0  (1 + \Theta \omega)}{\Theta \omega^2 + \omega + \Gamma_1}.
\end{equation}






Finally we solve for the time-dependence of the magnetization by taking the inverse Laplace transform,
\begin{equation}
m(t) = \mathcal{L}^{-1}[\frac{(1 + \Theta \omega)}{\Theta \omega^2 + \omega + \Gamma_1}] = \frac{\phi(t)(\Gamma_1 m_0 + \Gamma_2)}{\Gamma_1} - \frac{\Gamma_2}{\Gamma_1},
\end{equation}
we note that this bears a strong resemblance to the 
Markovian expression, Eq. \ref{m_v_t}, with the exponential being replaced by the function $\phi(t)$, which is 
\begin{equation}
\phi(t) = \frac{1}{2 \beta} \Big((\beta - 1) e^{-t(1+\beta)/2\Theta} + (\beta + 1) e^{-t(1-\beta)/2\Theta} \Big),
\label{phi}
\end{equation}
where  $\beta = \sqrt{1 - 4 \Gamma_1 \Theta}$. In the
limit $t \rightarrow \infty$, the value of the
magnetization again tends to $\frac{-\Gamma_2}{\Gamma_1}$.
To see that this agrees with the uncorrelated solution
for small correlation times, we may expand $\beta$ in $\Theta$ for
small $\Theta$, hence $\beta = 1 - 2 \Gamma_1 \Theta$, inserting into the magnetization it becomes
\begin{equation}
m(t) = \frac{\beta - 1}{2 \beta} e^{-t/2 \Theta} e ^{\Gamma_1t} + \frac{(\beta + 1)}{2\beta} e^{-\Gamma_1 t}.
\end{equation}

As $\Theta \rightarrow 0$, $\beta \rightarrow 1$, and only the second term in the expression for the magnetization
remains, $m(t) = e^{-\Gamma_1 t}$, so the small correlation time limit of the spin evolution agrees with the non Markovian
master equation.
\begin{figure}[ht]
\begin{center}
\includegraphics[width=0.5 \textwidth, height=\textheight,keepaspectratio]{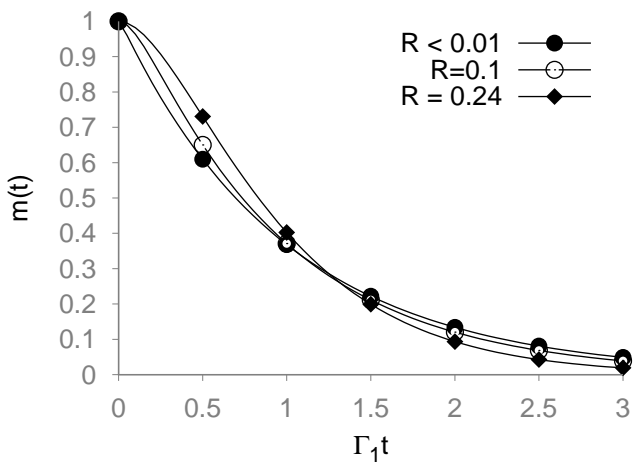}
\end{center}
\caption{$m(t)$ vs $t$, for $R=0,0.1,0.2$,
under the initial condition $m=1$, with
transition rates $\kappa_{12} = 1$, $\kappa_{21} = 0$}
\label{fig3}
\end{figure}

Finally, we note that the solution for the magnetization breaks down into two regimes. First, we note that the
expression for $\beta$  depends only on the product
of the correlation time, $\Theta$, and the rate $\Gamma_1$, and
not on their specific individual values. We may then
discuss the behavior of the model in terms of only the ratio
parameter $R = \Gamma_1 \Theta = \Theta/\Gamma_1^{-1}$, which gives the
ratio of the well correlation time to the escape time.
Rewriting the Eq.\ref{phi}  for the spin vs time,
\begin{eqnarray}
m(t) &=&  \frac{(\Gamma_1 m_0 + \Gamma_2)}{\Gamma_1} \Big( (e^{-t/2\Theta} ([e^{\beta t / {2\Theta}} \\ \nonumber &-& e^{-\beta t / {2\Theta}}]/2 \beta
+ [ e^{-\beta t/ 2 \Theta} + e^{\beta t/ 2 \Theta}]/2 ) \Big) - \frac{\Gamma_2}{\Gamma_1},
\end{eqnarray}
which may be simplified in terms of hyperbolic
trigonometric functions, 
\begin{equation}
m(t) = e^{-t/2\Theta} \Big( \frac{\sinh(\beta t / {2\Theta})}{\beta} + \cosh(\beta t / {2\Theta}) \Big). 
\label{m_v_t_cos}
\end{equation}

For smaller $R < \frac{1}{4}$, we have a real value of $\beta = \sqrt{1 - 4 R}$, and the time-dependence of the spin
corresponds to Eq. \ref{m_v_t_cos}. In Figure \ref{fig3}, we plot the time-evolution for values of $R < \frac{1}{4}$. Once
the correlation time is some sizable fraction of the escape time, the behavior begins to depart from the simple exponential behavior predicted in the Markovian system.
At early times the magnetization decays more slowly
than the exponential decay and at later times it decays more
quickly, 
while the timescale over which the decay occurs ($\Gamma_1$) remains the same. The effect of the increasing 
correlation time between the populations of the wells is then to shift the process to different, lower
frequencies.

In the case that $R > \frac{1}{4}$,  we have an imaginary argument to $\sinh$ and $\cosh$,
we then have an expression for $m(t)$

\begin{equation}
m(t) = e^{-t/2\Theta} ( \frac{\sin(b t / {2\Theta})}{b} + \cos(b t / {2\Theta}))
\end{equation}

where $b = \sqrt{4 R - 1}$. We note that
the solutions take the form of damped oscillations which tends toward the equilibrium value of the magnetization.
However, these solutions are unphysical as 
the occupation in individual wells may become
less than $0$ for these values.
This is not surprising, as for longer correlation times
 the generalized master equation will overestimate the population in each well and generate a time evolution which
 will continue to reduce the population of a well, even
 when that well is presently empty. It is also unclear
 what it would mean for the
 correlation time of the well population to exceed or
 be on the order of the overall escape time, as this would
 imply that the timescale over which the spin population is
 correlated exceeds the overall escape time for the system,
 which is itself determined by changes in the 
 individual well populations.

\section{Comparison}
We may now directly compare the magnetic relaxation profiles calculated from explicit numerical integration of Eqs.
\ref{LLMS-1}, \ref{LLMS-2} at various barrier heights, damping and correlation times, to the biexponential decay
predicted by the generalized master equation. In all of the present simulations we again use simulation parameters
comparable to the Co nanoparticle of volume $V = 8 \times 10^{-27} m^3$, anisotropy energy density
$K = 4.2 \times 10^5 J/m^3$ giving an anisotropy energy $KV = 1.12 \times 10^{21} J$, and a magnetic moment
$\mu_s = 1.12 \times 10^{-20} J/T$, while no external applied field is assumed, $H_{ext} = 0$. 
\begin{figure}[ht]
\begin{center}
\includegraphics[width=0.4 \textwidth, height=\textheight,keepaspectratio]{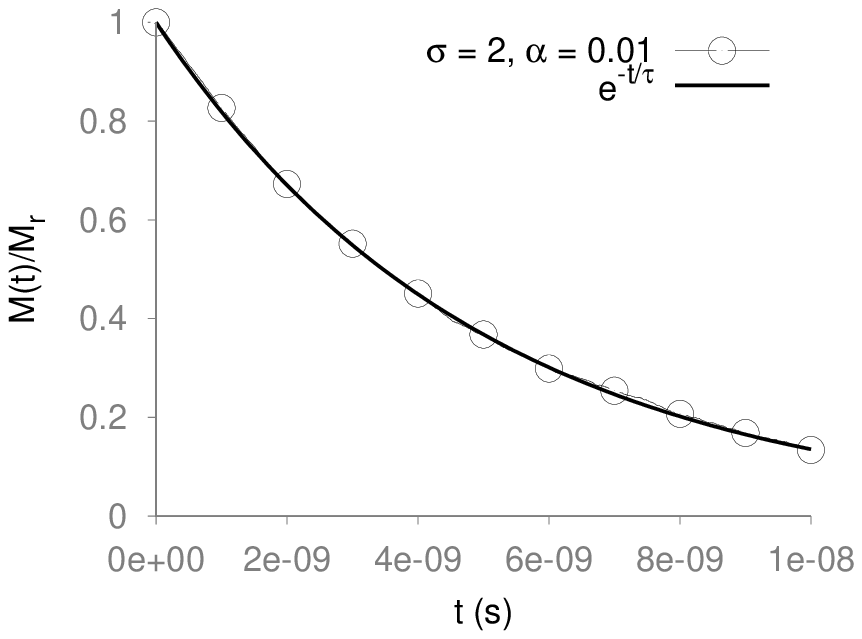}
\includegraphics[width=0.4 \textwidth, height=\textheight,keepaspectratio]{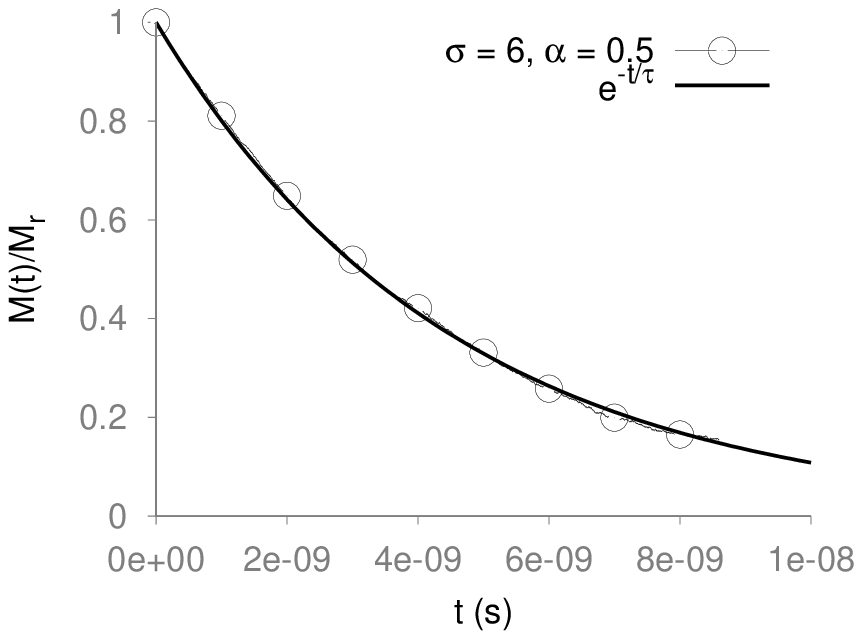}
\end{center}
\caption{Spin relaxation profiles from LLG simulations
for {\bf TOP} : $\sigma = 2$, $\alpha = 0.01$, giving an exponential decay with characteristic escape time $\tau = 5 \times 10^{-9}s$ and
{\bf BOTTOM} : $\sigma = 6$, $\alpha = 0.5$,  $\tau = 4.5 \times 10^{-9}s$}.
\label{fig5}
\end{figure}

The spins are initialized in the equilibrium Boltzmann distribution in one of the minima of the potential energy,
according to the distribution $P(\theta) \propto \sin(\theta) \exp (-k_u/k_B T \sin^2(\theta))$. To ensure that the noise
is equilibrated with the spin at the correct temperature, the noise is initially set to $\eta_{i,j,k} = 0$, and is then
evolved in the presence of the equilibrium distribution in the well until they come into thermal equilibrium. The initial 
condition of the noise is important, as, for example, a choice of $\eta(t=0) = 0$, will result in a field which
quickly aligns with the spins in the potential minimum and give an unphysical increase in the well population from
equilibrium at short times.

The time-evolution of the magnetization, $M(t) = \langle S_z(i) \rangle$ is then plotted, normalized by the initial remanent
magnetization inside of the well, $M_r = M(0)$.

\begin{figure}[ht]
\begin{center}
\includegraphics[width=0.4 \textwidth, height=\textheight,keepaspectratio]{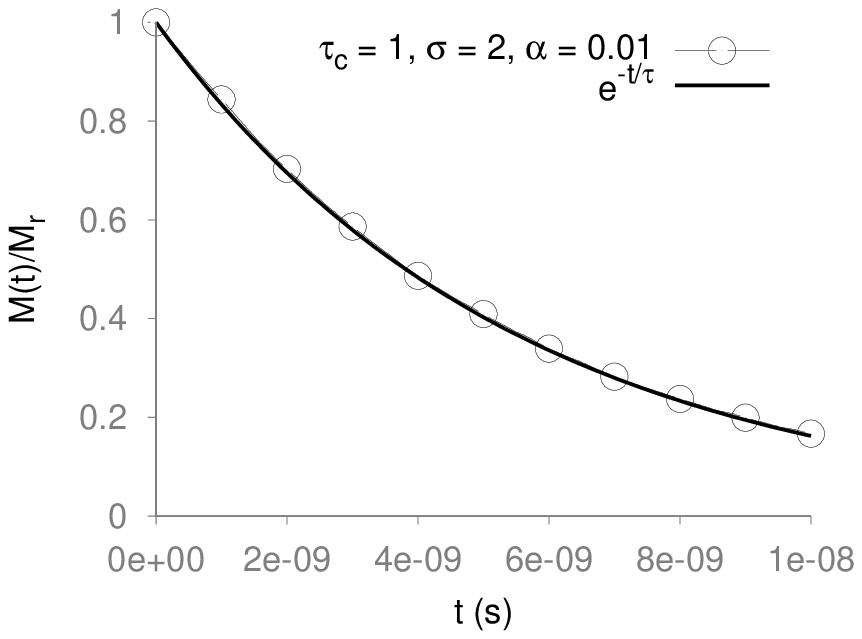}
\includegraphics[width=0.4 \textwidth, height=\textheight,keepaspectratio]{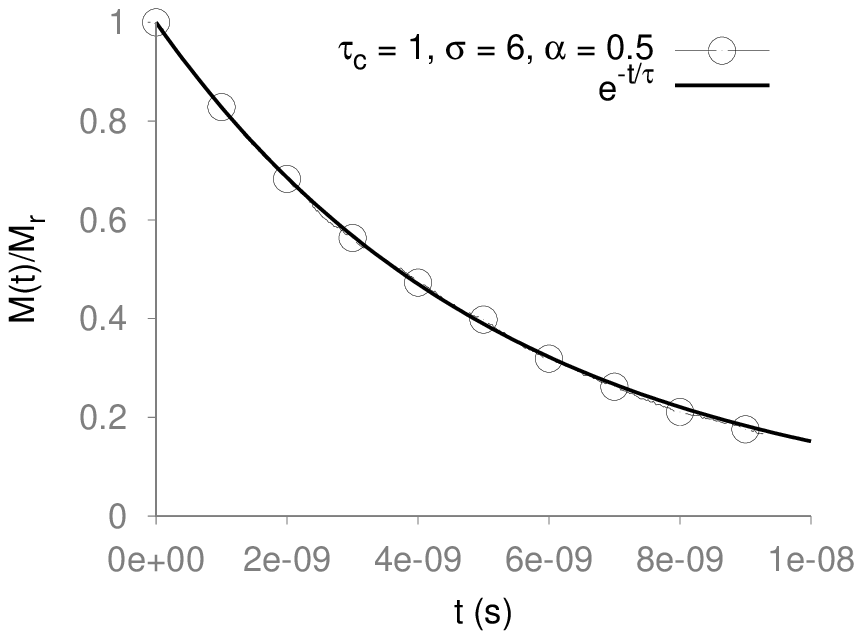}
\end{center}
\caption{Exponential behavior from LLMS simulations, for  {\bf TOP}: $\tau_c = 1$, $\sigma = 2$. $\alpha = 0.01$ we have an exponential decay with escape time $\tau = 5.5 \times 10^{-9}$, and {\bf BOTTOM}: $\tau_c = 1$, $\sigma = 6$. $\alpha = 0.5$ $\tau = 5.3 \times 0^{-9}$s
For low damping and large barrier heights, the correlation time is much smaller than the escape time.}
\label{fig6}
\end{figure}

In Figure \ref{fig5}, we depict the numerical calculation of the relaxation profile
from the LLG. This gives rise to an exponential behavior with a single relaxation time, 
which is directly comparable to the exponential decay of the master equation. In general,
the relaxation profile from the LLG may be non-exponential, with both the
integral relaxation time and the decay profile depending on the higher-order eigenvalues
of the Fokker-Planck operator and the equilibrium correlation functions of the spin, 
 $\tau^i_{int} = \sum_k \tau^i_k \lambda_k$. However, the relaxation is dominated
by the first eigenvalue in the high-barrier limit and for small applied fields
, for $\sigma > 1$, with good agreement
between the LLG and exponential decay for $\sigma$ as low as $2$, as is shown in Figure \ref{fig5}.

\begin{figure}[ht]
\begin{center}
\includegraphics[width=0.5 \textwidth, height=\textheight,keepaspectratio]{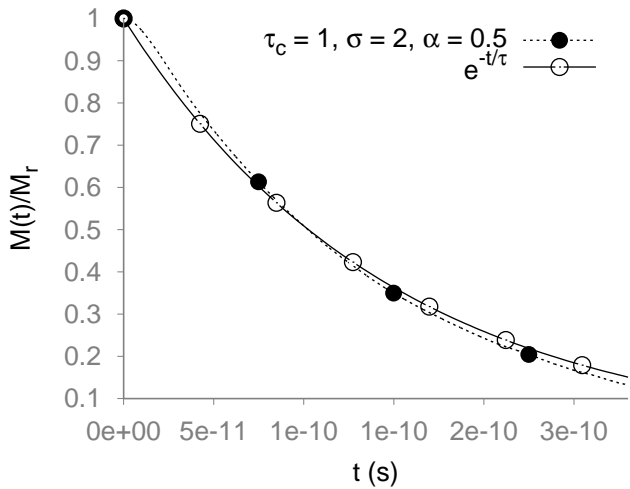}
\end{center}
\caption{Biexponential behavior from LLMS simulations for $\tau = 1$, $\sigma = 2$, $\alpha = 0.5$ and $\tau = 1.48 x 10^{-10}$}
\label{fig7}
\end{figure}

Figure \ref{fig6} shows the relaxation from LLMS simulations of the Co nanoparticle, where the correlation
time is chosen to be of the order of the inverse Larmor precession time such that $\tau_c \approx (\gamma H_k)^{-1}$.
In both the cases of low damping and higher barriers, we see that the ordinary exponential behavior of the
LLG is retained. In this case the escape time is much larger than the correlation time of the noise, and the
relaxational dynamics are unaffected by the intra-well dynamics of the spin which occur on a much
faster timescale than the relaxation, $\tau_c/\tau \approx 0.01$ for both simulations.

In the intermediate-to-high damping and high damping regimes, the behavior of the magnetization
becomes much more interesting and departs from the LLG. In particular, for a relatively small barrier of $\sigma = 2$,
$\alpha = 0.5$ and a correlation time again of the order of the inverse Larmor frequency. In this case the ratio of the
escape to the correlation time is $\tau_c/\tau = 9.4 \times 10^{-12}s/1.48 \times 10^{-10}s \approx 0.06$. The influence of the
spin correlation is now visible in the relaxation profile of the escape, as shown in Figure \ref{fig7}, which is similar to the
biexponential deviation predicted by the generalized master equation, with the relaxation proceeding more slowly
at earlier times and speeding up at later times.
\begin{figure}[ht]
\begin{center}
\includegraphics[width=0.4 \textwidth, height=\textheight,keepaspectratio]{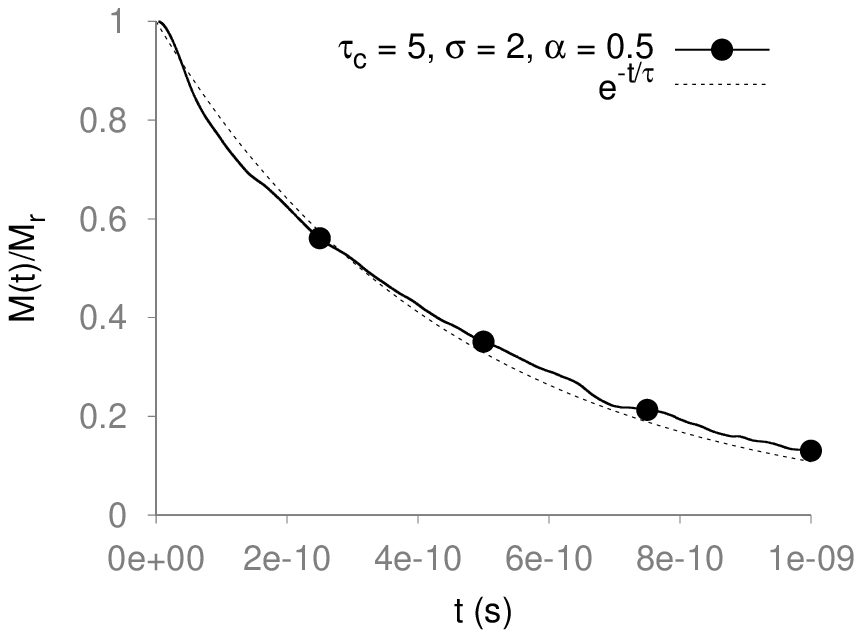}
\includegraphics[width=0.4 \textwidth, height=\textheight,keepaspectratio]{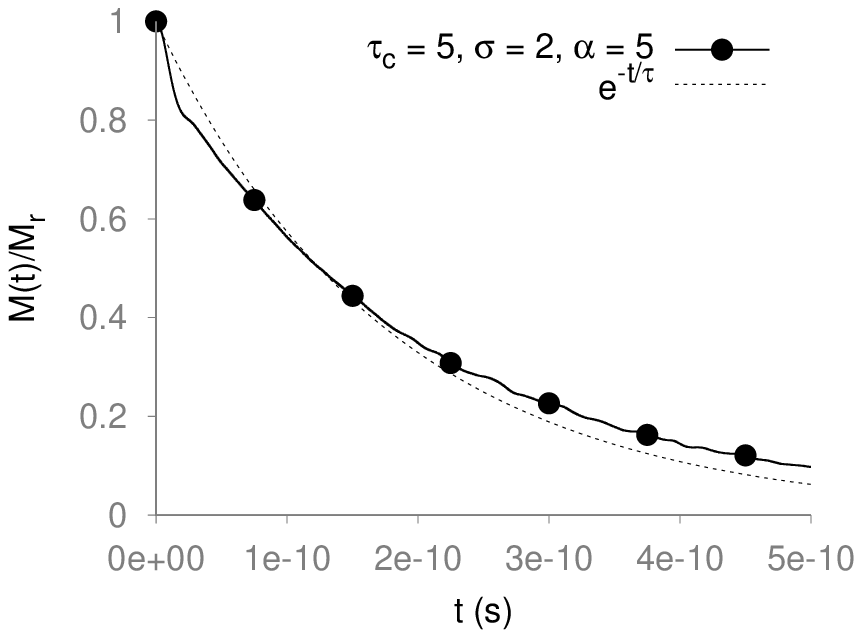}
\end{center}
\caption{LLMS simulations at high damping and long
correlation times. The behavior continues to depart
from a purely exponential decay, but now exhibits a noisy, more complicated time-dependence. 
{\bf TOP} : $\tau_c = 5$, $\sigma = 2$ , $\alpha = 0.5$ and $\tau = 4.5 \times 10^{-10}$, {\bf BOTTOM}:
$\tau_c = 5$, $\sigma = 2$ , $\alpha = 5$ and $\tau = 1.8 \times 10^{-10}$ }.
\label{fig8}
\end{figure}

Finally, for very long correlation times and high damping, the correlation time remains a sizable fraction of the
escape time. However the biexponential behavior is no longer evident as shown in figure \ref{fig8}. The decay
remains approximately exponential with a highly noisy path, a possible indication that the precise decay profile
is extremely dependent on the initial conditions for such strong coupling between the spin and bath.

\section{Conclusions}
We have investigated thermal relaxation in magnetic nanoparticles introducing colored noise. Two models are considered. The first is an approach based on the numerical solution of the Landau-Lifshitz-Miyazaki-Seki (LLMS)
model, which replaces the white noise approximation associated with the use of LLB-equation based models. Due to computational requirements the LLMS approach is useful for relatively short timescales, consequently a second approach is derived based on a generalized master equation approach involving the introduction of a memory kernel. We find that the importance of colored noise is determined by the ratio of the correlation time $\tau_c$ to the characteristic system time $\tau_s = (\gamma H_k)^{-1}$, which is essentially the Larmor precession time. Consequently correlated noise should become important for materials with large magnetic anisotropy such as SmCo\textsubscript{5} where the characteristic time approaches femtoseconds. Both models, the LLMS-based approach and the master equation, although derived for different timescales, exhibit an unusual bi-exponential decay of the magnetization, which represents an interesting signature of the presence of colored noise.

\appendix
\section{Colored Noise}
In this appendix we present some relevant background material on the LLMS equation and colored noise.

\subsection{Landau-Lifshitz-Miyazaki-Seki}
The LLMS equations constitute an implementation of a colored noise in a system with a thermalization condition
represented through the Fluctuation-Dissipation theorem. We reproduce here the original derivation by Miyazaki
and Seki \cite{3}, of the spin-only expression of the LLMS, which allows us to compare the LLMS thermal
fluctuations directly to the Ornstein-Uhlenbeck. The time evolution of the LLMS noise term is similar
to the OU, with an additional term which couples explicitly to the spin,
\begin{equation}
\frac{d \boldsymbol{\eta}}{dt} = -\frac{1}{\tau_c}\Big(\boldsymbol{\eta}(t) - \chi \mathbf{S}(t) \Big) + \mathbf{R}.
\end{equation}
Taking $D = \frac{\chi k_B T}{\mu_s}$, then the autocorrelation of the field $\mathbf{R}$ is 
$\langle \mathbf{R}(t) \mathbf{R}(t') \rangle = 2 \frac{D}{\tau_c} \delta(t - t')$, and proceeding
to solve as a first-order linear differential equation in the same manner as the OU noise, we have
\begin{eqnarray}
\boldsymbol{\eta}(t) = \frac{\chi}{\tau_c} \int_{-\infty}^t dt' & K & (t-t') \mathbf{S}(t') \\
& + & \sqrt{\frac{2D}{\tau_c}} \int_{-\infty}^t dt' K(t-t') \boldsymbol{\Gamma}(t). \nonumber
\end{eqnarray}
After integrating the first term by parts, we have
\begin{eqnarray}
\boldsymbol{\eta}(t) =  \sqrt{\frac{2D}{\tau_c}} \int_{-\infty}^t dt' & K & (t-t') \boldsymbol{\Gamma}(t) \\
& - & \chi \int_{-\infty}^t dt' K(t-t') \frac{d\mathbf{S}(t')}{dt'}, \nonumber
\end{eqnarray}
and by inserting this into the precessional equation for the spin, we get the spin-only form for the LLMS equation,
\begin{equation}
\frac{d\mathbf{S}}{dt} = \gamma \mathbf{S}(t) \times \Big(\mathbf{H} + \bar{\boldsymbol{\eta}} - \chi \int_{-\infty}^t dt' K(t-t') \frac{d\mathbf{S}(t')}{dt'}\Big),
\end{equation}
where we now label the thermal fluctuations by $\bar{\boldsymbol{\eta}}(t)$,
\begin{equation}
\bar{\boldsymbol{\eta}}(t) = \sqrt{\frac{2D}{\tau_c}} \int_{-\infty}^t dt' K(t-t') \boldsymbol{\Gamma}(t').
\end{equation}

The autocorrelation of this thermal field is
\begin{eqnarray}
\langle \bar{\boldsymbol{\eta}}(t) \bar{\boldsymbol{\eta}}(t') \rangle & = & D K(t-t') \\
& = & \frac{\chi k_B T}{\mu_s} K(t-t') = \frac{\beta^{-1}}{\mu_s} \chi K(t-t'), \nonumber
\end{eqnarray}
Recognizing $\chi K(t-t')$ as the damping term, we see that this is a representation
of the Fluctuation-Dissipation theorem for the colored noise, where the additional 
factor of $\mu_s$ arises from the spin normalization. Taking the zero correlation time limit,
\begin{equation}
\lim_{\tau_c \rightarrow 0} \langle \bar{\boldsymbol{\eta}}(t) \bar{\boldsymbol{\eta}}(t') \rangle = 2 D \tau_c \delta(t-t').
\end{equation}

\begin{figure}[ht]
\begin{center}
\includegraphics[width=0.4 \textwidth]{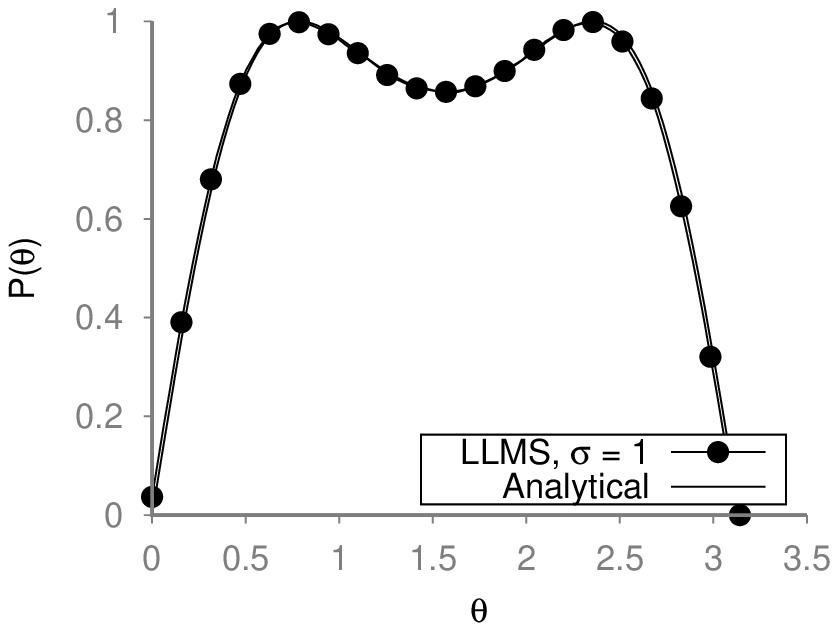}
\includegraphics[width=0.4 \textwidth]{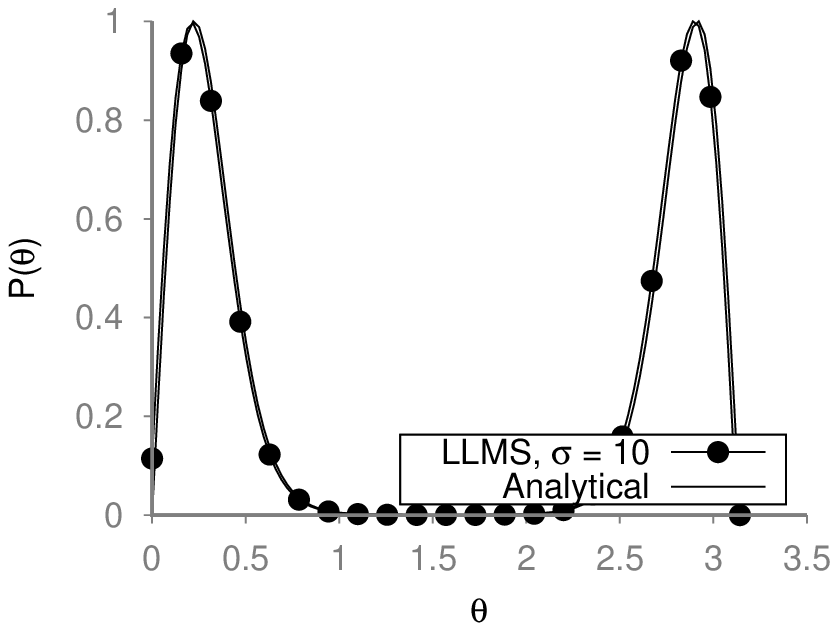}
\end{center}
\caption[Boltzmann Distribution from magnetic Langevin simulations.]{$P(\theta)$ vs $\theta$, from numerical simulations
of the LLMS equation for {\bf TOP}: $\sigma = 1$ and {\bf BOTTOM}: $\sigma = 10$, with $\tau_c \gamma H_k = 2$.}.
\label{boltzllg}
\end{figure}

We note that the LLMS thus derived from the physical consideration of the spin-field interaction is not
immediately comparable with the typical expression for the Ornstein-Uhlenbeck
colored noise, owing to the fact that the $1/\tau_c$ term has been implicitly
absorbed in the white noise term. If we rescale the driving noise such that $\mathbf{Q}(t) = \tau_c \mathbf{R}(t)$,
we then have a pair of Langevin equations
\begin{equation}
\frac{d\mathbf{S}}{dt} = \gamma (\mathbf{S} \times (\mathbf{H} + \boldsymbol{\eta})),
\end{equation}
while the noise evolves as,
\begin{equation}
\frac{d \boldsymbol{\eta}}{dt} = - \frac{1}{\tau_c} \Big(\boldsymbol{\eta}(t) - \chi \mathbf{S}(t) + \mathbf{Q} \Big).
\end{equation}
The autocorrelation of the white noise is
\begin{equation}
\langle \mathbf{Q}(t) \mathbf{Q}(t') \rangle = \frac{2 \chi \tau_c k_B T}{\mu_s} \delta(t-t') = 2 D \delta(t-t'),
\end{equation}
with $D = \frac{\chi \tau_c k_B T}{\mu_s}$, while the limit of the autocorrelation of the
thermal term in the spin-only expression is now,
\begin{equation}
\lim_{\tau_c \rightarrow 0} \langle \bar{\mathbf{Q}}(t) \bar{\mathbf{Q}}(t') \rangle = \frac{D }{\tau_c} \delta(t-t'),
\end{equation}
which is directly comparable to the Ornstein-Uhlenbeck form of the colored noise. The expression of the
LLMS in terms of the bath variable $Q$ has the additional benefit that $\big[\mathbf{Q}\big] = T$ and so
we can interpret $\mathbf{Q}$ as the thermal magnetic field contribution to the evolution of the
bath field.

Finally, we may see that the limit of the LLMS equation for vanishing correlation
time is the LLG equation.  For small correlation times we can then take the Taylor expansion about the time $t$ in $t'$, so that
the damping term becomes,
\begin{equation}
\int_{-\infty}^t K(t-t') \frac{d\mathbf{S}(t')}{dt'}dt' =  \Big[ \int_{-\infty}^t K(t') dt' \Big] \frac{d\mathbf{S(t)}}{dt} + ...
\end{equation}
Hence the spin and memory kernel decouple in the small correlation time limit, and the Langevin equation
becomes
\begin{equation}
\frac{d\mathbf{S}}{dt} = \gamma \mathbf{S}(t) \times \Big(\mathbf{H} + \bar{\boldsymbol{\eta}} - \Big[ \chi \int_{-\infty}^t dt' K(t-t') \Big] \frac{d\mathbf{S}(t)}{dt}\Big),
\end{equation}
After performing the integration over $t'$, the damping is
\begin{equation}
\chi \int_{-\infty}^t dt' e^{-(t-t')/\tau_c} = \chi \tau_c .
\end{equation}
and by direct comparison of the damping terms in this expression and in Gilbert's equation we have the relationship
of the phenomenological damping to the LLMS parameters $\alpha = \chi \gamma \tau_c$. We note also that this expression 
can be seen if we identify the driving white noise in the bath field of the LLMS with the thermal magnetic fields of the LLG.
\begin{equation}
\begin{split}
\langle \mathbf{Q}(t) \mathbf{Q}(t') \rangle = \frac{2 \chi \tau_c k_B T}{\mu_s} \delta(t-t') 
\\ = \frac{2 \alpha k_B T}{\gamma \mu_s} \delta(t-t') 
\\ = \langle \mathbf{H}_{th}(t) \mathbf{H}_{th}	(t') \rangle
\end{split}
\end{equation}
under the assumption that $\alpha = \gamma \chi \tau_c$. 

\subsection{Thermalization}
As a quantitative evaluation of the LLMS model and our implementation thereof, we compare the equilibrium behavior to the appropriate analytical Boltzmann distribution, which the Markovian LLG equation also satisfies. We simulate a single spin under the influence 
of anisotropy only. The Boltzmann distribution for such a system is 
\begin{equation}
P(\theta) \propto \sin \theta \exp (\frac{-k_u \sin^2 \theta}{k_B T})
\end{equation}
where $\theta$ is the angle between the spin and the easy-axis and the factor of $\sin \theta$ arises from 
normalizing the probability distribution on the sphere. WE initialize the spin along the easy-axis direction, 
then allow the spin to evolve for $10^8$ steps after equilibration and evaluate the probability distribution
by recording the number of steps the spin spends at 
each angle to the easy-axis.

In Figure \ref{boltzllg}, we compare the numerical results to the analytical expression for
both the LLMS model and the standard LLG augmented by
Ornstein-Uhlenbeck fields of the type generated by
the Langevin equation in Eq. \ref{Langevin}. The simulations
using the LLMS model agree with the anticipated Boltzmann
distribution at equilibrium, while the LLG with
Ornstein-Uhlenbeck fails to reproduce the correct distribution. This is because, as we have argued, this does not comprise a correct implementation of the Fluctuation-Dissipation theorem, with deviations corresponding to the missing high-frequency components of the damping.

\vspace{15mm}

\bibliographystyle{abbrv}

\end{document}